\documentclass[twocolumn,
               showpacs,
               preprintnumbers,
               prl]{revtex4}
\usepackage{graphicx,color}

\usepackage{ulem}
\usepackage{amssymb}
\usepackage{amsbsy}
\usepackage{amsthm,mathrsfs,amsopn}
\usepackage{amsmath}

\theoremstyle{plain} 

\newtheorem{theorem}{Theorem}[section]

\newtheorem{remark}[theorem]{Remark}

\begin{document}

\title{Turing patterns in multiplex networks}

\author{Malbor Asllani$^{1,2}$, Daniel M. Busiello$^{2}$, Timoteo Carletti$^{3}$, Duccio Fanelli$^{2}$, Gwendoline Planchon$^{2,3}$}
\affiliation{
1. Dipartimento di  Scienza e Alta Tecnologia, Universit\`{a} degli Studi dell'Insubria, via Valleggio 11, 22100 Como, Italy\\
2. Dipartimento di Fisica e Astronomia, University of Florence,INFN and CSDC, Via Sansone 1, 50019 Sesto Fiorentino, Florence, Italy \\
3. Department of Mathematics and Namur Center for Complex Systems - naXys, University of Namur, rempart de la Vierge 8, B 5000 Namur, Belgium}

\begin{abstract} 
The theory of patterns formation for a reaction-diffusion system defined on a multiplex is developed by means of a perturbative approach. The intra-layer diffusion constants act as small parameter in the expansion and the unperturbed state coincides with the limiting setting where the multiplex layers are decoupled. The interaction between adjacent layers can seed the instability of an homogeneous fixed point, yielding 
self-organized patterns which are instead impeded in the limit of decoupled layers. Patterns on individual layers can also fade away due to cross-talking between layers. Analytical results are compared to direct simulations.   
\end{abstract}

\maketitle

Patterns are widespread in nature: regular forms and geometries, like spirals, trees and stripes, recur in different contexts. Animals present magnificient and colorful patterns \cite{murray}, which often call for evolutionary explainations. Camouflage and signalling are among the functions that patterns exert, acting as key mediators of animal behaviour and sociality. Spatial motifs emerge in stirred chemical reactors \cite{zhab}, exemplifying a spontaneous drive for self-organization which universally permeate{s} life in all its manifestations, from cells to large organism, or communities. In a seminal paper Alan Turing set forth a theory by which patterns formation might arise from the dynamical interplay between reaction and diffusion in a chemical system \cite{turing}. Turing ideas provide a plausible and general explaination of how a variety of patterns can emerge in living systems. Under specific condition{s}, diffusion drives an instability by perturbing an homogeneous stable fixed point, via an activator-inhibitor mechanism. As the perturbation grows, non linear reactions balance the diffusion terms, 
yielding the asymptotic, spatially inhomogeneous, steady state.  Usually, reaction diffusion models are defined on a regular
lattice, either continuous or discrete. In many cases of interest, it is however more natural to schematize the system as a complex network.
With reference to ecology, the nodes of the networks mimics localized habitat patches, and
the dispersal connection among habitats result in the diffusive coupling between adjacent
nodes. In the  brain a network of neuronal connections is active, which
provide the backbone for the propagation of the cortical activity. The internet and the
cyberword in general are other, quite obvious examples that require invoking
the concept of network. Building on the pionering work of Othmer and Scriven \cite{othmer}, Nakao and Mikhailov developed in \cite{nakao} the theory of Turing patterns formation on random symmetric network, highlighting the peculiarities
that stem from the embedding graph structure. More recently, the case of directed, hence non symmetric,
networks has been addressed \cite{asllaniNatureComm}. When the reactants can only diffuse along allowed routes, the tracks that correspond to the
reversal moves being formally impeded,  topology driven instabilities can develop also when the system under scrutiny 
cannot experience a Turing like (or wave instability) if defined on a regular lattice or, equivalently, on a continuous spatial support.

\begin{figure}[h]
\begin{center}
\hspace*{-1cm}
\includegraphics[scale=0.3]{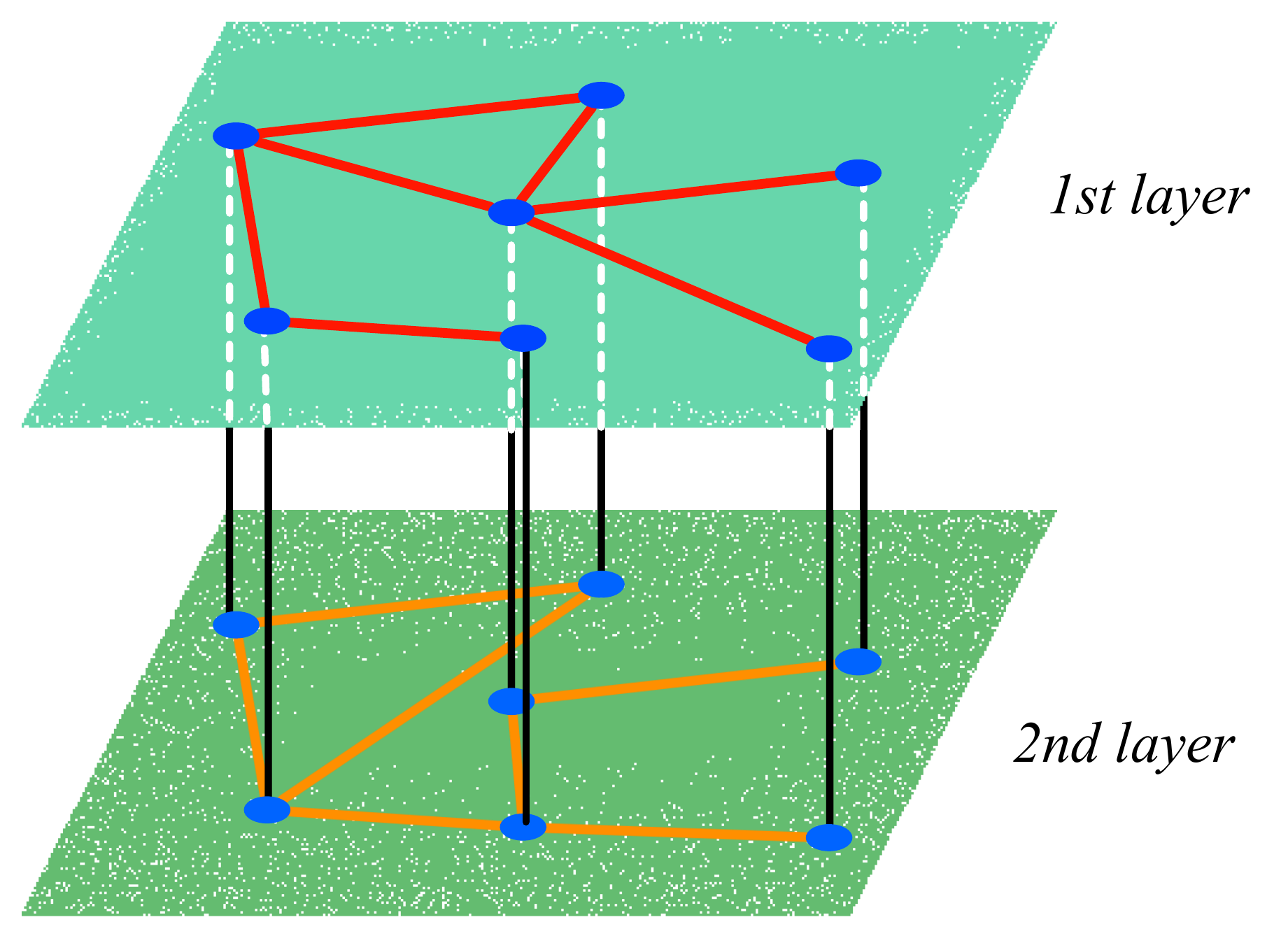}
\end{center}
\caption{A schematic illustration of a two layers multiplex network.}
\label{fig:Fig0}
\end{figure}

However, the conventional approach to network theory is not general enough to ascertain the complexity that hides 
behind real world applications. Self-organization may proceed across multiple, inter-linked networks, by exploiting the multifaceted nature of resources and organizational skills. For this reason, multiplex, networks in layers whose mutual connections are between twin nodes, see Figure \ref{fig:Fig0}, have been introduced as a necessary leap forward in the modeling effort \cite{multi1,multi2,multi3, multi4, multi5}. These concepts are particularly relevant to transportation systems \cite{kuran,zou}, the learning organization in the brain \cite{brain} and to understanding the emergent dynamics in social commmunities \cite{social}. In \cite{arenas} the process of single species diffusion on a multiplex networks has been investigated, and the spectrum of the associated Laplacian matrix characterized in term of its intra- and interlayer structure. 

In this Letter we build on these premises to derive a general theory of patterns formation for multispecies reaction diffusion systems on a multiplex. Cooperative interference between adjacent layers manifests, yielding stratified patterns also when the Turing like instability on each individual layer is impeded. Conversely, patterns can dissolve as a consequence of the intra-layer overlap. The analysis is carried out analytically via a perturbative scheme which enables to derived closed analytical expressions for the critical coupling {that} determines the aforementioned transitions. The adequacy of the analytical predictions is confirmed by direct numerical simulations. 

We begin the discussion by reviewing the theory of Turing patterns on a monolayer network made of $\Omega$ nodes and characterized by the $\Omega \times \Omega$ adjacency matrix $\mathbf{W}$. $W_{ij}$ is equal to one if nodes $i$ and $j$ (with $i \ne j$) are connected, and zero otherwise. 
We here consider undirected networks, which implies that the matrix $\mathbf{W}$ is symmetric. A two species reaction diffusion system can be cast in the general form:

\begin{eqnarray}
\frac{d u_i}{d t} &=& f(u_i,v_i)+ D_{u}\sum_j L_{ij} u_j \nonumber\\
\frac{d v_i}{d t} &=& g(u_i,v_i)+D_{v} \sum_j L_{ij} v_j
\label{eq:reac_dif}
\end{eqnarray}
where $u_i$ and $v_i$ stand for the concentrations of the species on node $i$. $L_{ij}=W_{ij}-k_i \delta_{ij}$ is the network Laplacian, where $k_i=\sum_j W_{ij}$ refers to the connectivity of node $i$ and $\delta_{ij}$ is the Kronecker's delta. $D_{u}$ and $D_{v}$ denote the diffusion coefficients; $f(\cdot, \cdot)$ and $g(\cdot, \cdot)$ are nonlinear functions of the concentrations and specify 
the reaction dynamics of the activator, which autocatalytically enhances its own production, and of the inhibitor, which contrast in turn the activator growth. Imagine that system (\ref{eq:reac_dif}) admits an homogeneous fixed point, $(\hat{u},\hat{v})$. This amounts to require $f(\hat{u},\hat{v})=g(\hat{u},\hat{v})=0$. Assume also that $(\hat{u},\hat{v})$ is stable, i.e. $\textrm{tr}(\textbf{J})=f_{u}+g_{v}<0$ and $\textrm{det}(\textbf{J})=f_{u} g_{v}-f_{v} g_{u}>0$, where $\textbf{J}$ is the Jacobian matrix associated to system (\ref{eq:reac_dif}).  As usual $f_u$, $f_v$, $g_u$ and $g_v$ stands for the partial derivatives of the reaction terms, evaluated at the equilibrium point $(\hat{u},\hat{v})$. Patterns (waves) arise when $(\hat{u},\hat{v})$ becomes unstable with respect to inhomogeneous perturbations. To look for instabilities, one can introduce a small perturbation ($\delta u_i$, $\delta v_i$) to the fixed point and linearize around it. In formulae:
\begin{equation} \label{eq:linear}
	\begin{pmatrix} \delta \dot u_i \\ \delta \dot v_i \end{pmatrix} = \sum_{j=1}^{\Omega} \left( \textbf{J} \delta_{ij} +   \textbf{D} L_{ij} \right) \begin{pmatrix} \delta u_j \\ \delta v_j \end{pmatrix},
\end{equation}
where $\textbf{D}= \left( \begin{smallmatrix} D_{u}&0\\ 0&D_{v} \end{smallmatrix} \right)$.

Following \cite{nakao} we introduce the eigenvalues and eigenvectors of the Laplacian operator $\sum_{j=1}^{\Omega} L_{ij} \Phi^{(\alpha)}_j =  \Lambda^{(\alpha)} \Phi^{(\alpha)}_i, \quad \alpha = 1,\ldots,\Omega$ and expand \footnote{Since the network is undirected, and the Laplacian operator is symmetric,  the eigenvalues $\Lambda^{(\alpha)}$ are real (and negative) and the eigenvectors $\Phi^{(\alpha)}$ form an orthonormal basis.} the inhomogeneous perturbations $\delta u_i$ and $\delta v_i$ as 
$\delta u_i{(t)}=\sum_{\alpha=1}^{\Omega}c_{\alpha}e^{\lambda_\alpha t}\Phi^{(\alpha)}_i$ and 
$\delta v_i{(t)}=\sum_{\alpha=1}^{\Omega}b_{\alpha}e^{\lambda_\alpha t}\Phi^{(\alpha)}_i$.
The constants $c_\alpha$ and $b_\alpha$ depend on the initial conditions. By inserting the above expressions in  
Eq.~\eqref{eq:linear} one obtains $\Omega$ independent linear equations for each different normal mode, yielding  
the eigenvalue problem $\textrm{det} \left( \textbf{J}_{\alpha}-\textbf{I}\lambda_{\alpha} \right)=0$,
where $\textbf{J}_{\alpha} \equiv \textbf{J}+\textbf{D}\Lambda^{(\alpha)}$ and $\textbf{I}$ stands for the $2 \times 2$ identity matrix.
The eigenvalue with the largest real part, defines the so-called dispersion relation
and characterizes the response of the system (\ref{eq:reac_dif}) to external perturbations.  If the real part of $\lambda_{\alpha}\equiv\lambda(\Lambda^{(\alpha)})$ is positive the initial perturbation grows exponentially in the linear regime of the evolution. Then, non linear effects become important and the system settles down into a non homogenoeus stationary configuration, characterized by a spontaneous polarization into activators-rich and inhibitors-poor groups. From hereon we assume  $\lambda_{\alpha}$ to label the (real) dispersion relation.    

Let us now turn to considering the reaction diffusion dynamics on a multiplex composed by two distinct layers. The
analysis readily extends to an arbitrary number of independent layers. For the sake of simplicity we will here
assume each layer to be characterized by an identical set of $\Omega$ nodes; the associated connectivity can however differ on each layer, as specified by the corresponding adjacency matrix  $W_{ij}^K$, with $i,j=1, \dots,\Omega$ and $K=1,2$. In principle the adjacency matrix can be weighted.  The species concentrations are denoted {by} $u_i^K$ and $v_i^K$ where the index $K$ identifies the layer to which the individuals belong. Species are allowed to diffuse on each layer, moving towards adjacent nodes with diffusion constants respectively given by  $D_u^K$ and $D_v^K$. Intra-layer diffusion is also accommodated for, via Fickean contributions which scale as the local concentration gradient, $D_u^{12}$ and $D_v^{12}$ being the associated diffusion constants. We hypothesize that reactions take place between individuals sharing the same node $i$ and layer $K$, and are formally coded via the non linear functions $f(u^K_i,v^K_i)$ and $g(u^K_i,v^K_i)$. Mathematically, the reaction-diffusion scheme (\ref{eq:reac_dif}) generalizes to:
\begin{equation}
\begin{cases}
\dot{u}^K_i&= f(u^K_i,v^K_i) + D_u^K\sum_{j=1}^{\Omega}L_{ij}^K u^K_j + D_u^{12} \left(u^{K+1}_i-u^K_i\right) \\
\dot{v}^K_i&= g(u^K_i,v^K_i) + D_v^K\sum_{j=1}^{\Omega}L_{ij}^K v^K_j + D_v^{12} \left(v^{K+1}_i-v^K_i\right) \\
\end{cases}
\label{eq:reac_diff}
\end{equation}
with $K=1,2$ and assuming $K+1$ to be $1$ for $K=2$. Here $L_{ij}^K=W_{ij}^K-k^K_i\delta_{ij}$ stands for the Laplacian matrix on the layer $K$. If the intra-layer diffusion is silenced, which implies setting $D_u^{12}=D_v^{12}=0$, the layers are decoupled. Working in this limit, one recovers hence two independent pair{s} of coupled reaction diffusion equation{s} for, respectively, $(u^1_i,v^1_i)$ and  $(u^2_i,v^2_i)$. Turing patterns can eventually set in for each of the considered limiting reaction-diffusion system as dictated by their associated dispersion relations  $\lambda_{\alpha_K}^K \equiv\lambda(\Lambda^{(\alpha_K)})$ with $K=1,2$, derived following the procedure outlined above. We are here instead  interested in the general setting where the inter-layed diffusion is accounted for. Can the system develop self-organized patterns which result from a positive interference between adjacent layers, when the instability is prevented to occur on each isolated level? Conversely, can patterns fade away when the diffusion between layers is {switched} on?         

To answer to these questions we adapt the above linear stability analysis to the present context. Linearizing around the stable homogeneous fixed point ($\hat{u}$,  $\hat{v}$) returns: 
\begin{equation}
\left( \begin{array}{ccc}
\dot{\delta\boldsymbol{u}}\\\dot{\delta\boldsymbol{v}}
 \end{array} \right)=\boldsymbol{\mathcal{\tilde{J}}}\left( \begin{array}{ccc}
\delta\boldsymbol{u}\\\delta\boldsymbol{v}
 \end{array} \right)
 \label{eq:lin_prob}
\end{equation}
with
\begin{equation*}
\boldsymbol{\mathcal{\tilde{J}}}=\left( \begin{array}{ccc}
f_u \mathbf{I}_{2\Omega} + \boldsymbol{\mathcal{L}}_u +D_u^{12}\boldsymbol{\mathcal{I}} & f_v \mathbf{I}_{2\Omega}\\
g_u \mathbf{I}_{2\Omega} & g_v \mathbf{I}_{2\Omega} + \boldsymbol{\mathcal{L}}_v+D_v^{12}\boldsymbol{\mathcal{I}}
 \end{array} \right)
 \label{eq:lin_prob2}
\end{equation*}
and where we have introduced the compact vector notation $\boldsymbol{x}=\left(x^1_1,\dots,x^1_{\Omega},x^2_1,\dots,x^2_{\Omega}\right)^T$, for $x=u,v$. Also, $\boldsymbol{\mathcal{I}}=\left(\begin{smallmatrix} -\textbf{I}_{\Omega} & \textbf{I}_{\Omega}\\ \textbf{I}_{\Omega} & -\textbf{I}_{\Omega}\end{smallmatrix}\right)$, where $\textbf{I}_{\Omega}$ denotes the $\Omega \times \Omega$-identity matrix. The multiplex Laplacian for the species $u$ reads:
$\boldsymbol{\mathcal{L}}_u=\left( \begin{array}{ccc}
D_u^1\mathbf{L}^1 & \mathbf{0}\\
\mathbf{0} & D_u^2\mathbf{L}^2
 \end{array}\right)\,$.
A similar operator, $\boldsymbol{\mathcal{L}}_v$, is associated to species $v$. Notice that   
$\boldsymbol{\mathcal{L}}_u +D_u^{12}\boldsymbol{\mathcal{I}}$ is the supra-Laplacian introduced in \cite{arenas}. Analogous consideration holds for the term that controls the migration of $v$ across the multiplex. Studying the $4 \Omega$ eigenvalues $\lambda$ of matrix $\boldsymbol{\mathcal{\tilde{J}}}$ ultimately returns the condition for the dynamical instability which anticipates  the emergence of Turing like patterns. If the real part of at least one {of} the $\lambda_i$, with $i=1,...,4 \Omega$ is positive,  the initial perturbation grows exponentially in the linear regime of the evolution. Non linear effects become then important and the system eventually attains a non homogenoeus stationary configuration. Unfortunately, in the multiplex version of the linear calculation, 
and for a generic choice of the diffusion constants, one cannot introduce a basis to expand the perturbations which 
diagonalizes the supra-Laplacian operators. In practice, one cannot project the full $4 \Omega \times 4 \Omega$ eigenvalue problem into a subspace of reduced dimensionality, as it is instead the case when the problem is defined on a single layer.
Moreover, it is not possible to exactly relate the spectrum of the multiplex matrix $\boldsymbol{\mathcal{\tilde{J}}}$ to those obtained when the layers are decoupled. Analytical insight can be gained through an apt perturbative algorithm which enables us to trace the modifications on the dispersion relation, as due to the diffusive coupling among layers. To this end we work in the limit of a weakly coupled multiplex, the inter-diffusion constants being instead assumed order one. Without losing generality we set $\epsilon \equiv D_v^{12} << 1$, and assume $D_u^{12}$ to be at most $O(\epsilon)$. We hence write $\boldsymbol{\mathcal{\tilde{J}}}=\boldsymbol{\mathcal{\tilde{J}}}_0+\epsilon\boldsymbol{\mathcal{D}}_0$ where 
$\boldsymbol{\mathcal{\tilde{J}}}_0=\left( \begin{array}{ccc}
f_u \mathbf{I}_{2\Omega} + \boldsymbol{\mathcal{L}}_u & f_v \mathbf{I}_{2\Omega}\\
g_u \mathbf{I}_{2\Omega} & g_v \mathbf{I}_{2\Omega} + \boldsymbol{\mathcal{L}}_v
 \end{array} \right)$ and $\boldsymbol{\mathcal{D}}_0 = \left( \begin{array}{ccc}
\frac{D_u^{12}}{D_v^{12}}\boldsymbol{L}^1 & \mathbf{0}\\
\mathbf{0} & \boldsymbol{L}^2
 \end{array} \right)$ .

The spectrum of $\boldsymbol{\mathcal{\tilde{J}}}_0$ is obtained as the union of the spectra of the two sub-matrices which define the condition for the instability on each of the layers taken independently. To study the deformation of the spectra produced by a small positive perturbation $\epsilon$, we refer to a straightforward extension of the Bauer-Fike theorem~\cite{golub}. 
We here give a general derivation of the result which will be then exploited with reference to the specific problem under investigation. 
Consider a  matrix $A_0$ under the assumption that the eigenvalues of $A_0$, $(\lambda_m^{(0)})_m$, have all multiplicity $1$. The associated eigenvectors, $(\mathbf{v}_m^{(0)})_m$ are thus linearly independent and form a basis for the underlying vector space $\mathbb{R}^{\Omega}$ (or $\mathbb{C}^{\Omega}$). Introduce now $A=A_0+\epsilon A_1$, $A_1$ representing the pertubation rescaled by $\epsilon$. We will denote with $\lambda(\epsilon)$ and $(\mathbf{v}_m(\epsilon))_m$  the eigenvalues and eigenvectors of matrix $A$. Let us introduce the matrices $\Lambda(\epsilon)=diag(\lambda_{1}(\epsilon),\lambda_{2}(\epsilon), \ldots \lambda_{\Omega}(\epsilon))$  and $V(\epsilon)=\left(\begin{matrix} \mathbf{v}_1(\epsilon)&
\mathbf{v}_2(\epsilon) &
\ldots &  \mathbf{v}_{\Omega}(\epsilon) \end{matrix}\right)$ and 
expand them into power of $\epsilon$ as:
\begin{equation}
\label{ps}
\Lambda(\epsilon)=\sum_{l\geq 0} \Lambda_l \epsilon^l \quad\text{and}\quad V(\epsilon)=\sum_{l\geq 0} V_l \epsilon^l\, ,
\end{equation}
where $\Lambda_0$ stands for the eigenvalues of the unperturbed matrix; $V_0$ (resp. $U_0$, to be used later) stands for the matrix whose columns (resp. rows) are the right (resp. left) eigenvectors of  $\boldsymbol{\mathcal{\tilde{J}}}_0$. Inserting formulae (\ref{ps}) into the perturbed system $(A_0+\epsilon A_1)V=V\Lambda$ and collecting together the terms of same order in $\epsilon$ beyond the trivial zero{-th} order contribution, we get
$A_0V_l+A_1V_{l-1}=\sum_{k=0}^lV_{l-k}\Lambda_k{\quad\forall l\geq 1}$.
{Left} mutiplying the previous equation by $U_0$ and setting $C_l=U_0 V_l$ yields:
\begin{equation}
\label{eqCLambda}
\Lambda_0C_l-C_l\Lambda_0=-U_0A_1V_{l-1}+C_0\Lambda_l+\sum_{k=1}^{l-1}C_{l-k}\Lambda_k\, .
\end{equation}
which can be solved (see Supplementary Material) to give $(\Lambda_l)_{ii} = (U_0 A_1V_{l-1})_{ii}$ ($(\Lambda_l)_{ij}=0$ for $i \ne j$) and $(C_l)_{ij} = 
\frac{(-U_0 A_1V_{l-1})_{ij}+\sum_{k=1}^{l-1}(C_{l-k}\Lambda_k)_{ij}}{\lambda^{(0)}_i-\lambda^{(0)}_j}$ ($(C_l)_{ii}=0$).

The above expressions allows us to asses the effect of the intra-layer coupling on the stability of the system. Select the eigenvalue with
the largest real part $\lambda_0^{max}$ of the unperturbed matrices $\boldsymbol{{\tilde{J}}}_0$ . For sufficiently small $\epsilon$, such that the relative ranking of the eigenvalues is preserved, we have at the leading order correction:   
\begin{equation}
{\lambda}^{max} (\epsilon) = \lambda_0^{max}+\epsilon  \frac{(U_0\boldsymbol{\mathcal{D}}_0V_0)_{kk}}{(U_0V_0)_{kk}}+\mathcal{O}(\epsilon^2)\, ,
\label{eq:pert_lambda}
\end{equation}
where $k$ is the index which refer to the largest unperturbed eigenvalue $\lambda_0^{max}$. Higher order corrections can be also computed as follows the general procedure outlined above. To illustrate how intra-layers couplings interfere with the ability of the system to self-organize in collective patterns, we apply the above analysis to a specific case study, the Brusselator model. This is a two species reaction-diffusion model whose local reaction terms are given by $f(u,v)=1-(b+1) u+cu^2 v $ and $g(u,v)=b u-cu^2 v $, where $b$ and $c$ act as constant parameters.

\begin{figure}[ht]
\begin{center}
\hspace*{-1cm}
\includegraphics[scale=0.3]{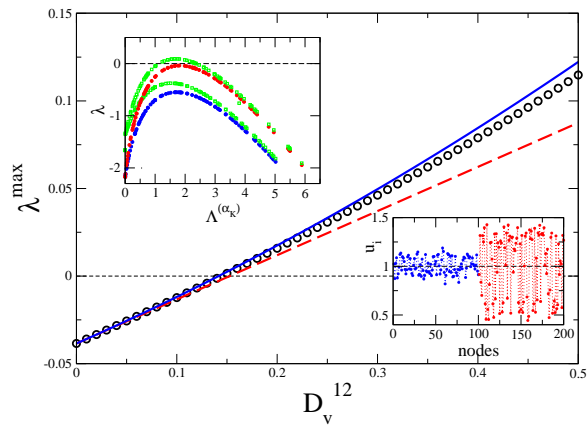}\\
\end{center}
\caption{Main: $\lambda^{max}$ is plotted versus $D_v^{12}$, starting from a condition for which the instability cannot occur when {$D_v^{12}=0$}. Circles refer to a direct numerical {computation} of $\lambda^{max}$. The dashed (resp. solid) line represents the analytical solution as obtained at the first (resp. second) perturbative order. Upper inset: the dispersion relation $\lambda$  is plotted versus the  eigenvalues of the (single layer) Laplacian operators, $L^1$ and $L^2$. The circles (resp. red and blue online) stand for $D_u^{12}=D_v^{12}=0$, while the squares (green online) are analytically calculated from (\ref{ps}), at the second order, for $D_u^{12}=0$ and $D_v^{12}=0.5$. The two layers of the multiplex have been generated as Watts-Strogatz (WD) \cite{WS} networks with probability of rewiring $p$ respectively equal to  $0.4$ and $0.6$.  The parameters are  $b=8, c=17, D^1_u=D^2_u=1, D^1_v=4, D^2_v=5$. Lower inset: asymptotic concentration of species $u$ as function of the nodes index $i$. The first (blue online) $\Omega=100$ nodes refer to the network with $p=0.4$, the other $\Omega$ (red online) to $p=0.6$.
 }
\label{fig:Fig1}
\end{figure}

Suppose now that for $\epsilon=0$ the system is stable, namely that $\lambda_0^{max}<0$, as depicted in the main panel of Figure \ref{fig:Fig1}.
No patterns can hence develop on any of the networks that define the layers of the multiplex. For an appropriate choice of the parameters of the model, $\lambda^{max}$ grows as function of the intra-layer diffusion $D_v^{12}$ ($=\epsilon$) and becomes eventually positive, signaling the presence of an instability which is specifically sensitive to the multiplex topology. The circles in Figure~\ref{fig:Fig1}  are computed by numerically calculating the eigenvalues of the matrix $\boldsymbol{\mathcal{\tilde{J}}}$ for different choices of the diffusion constant  $D_v^{12}$. The dashed line refer to the linear approximation (\ref{eq:pert_lambda}) and returns a quite reasonable estimate for the critical value of the intra-layer diffusion $D_{v,{crit}}^{12}$ for which the multiplex instability sets in, $D_{v,{crit}}^{12} \simeq - \lambda_0^{max} (U_0V_0)_{kk} / (U_0\boldsymbol{\mathcal{D}}_0V_0)_{kk} $. The solid line is obtained by accounting for the next-to-leading corrections in the perturbative calculation. In the upper inset of Figure~\ref{fig:Fig1} the dispersion relation is plotted versus $\Lambda^K_{\alpha_K}$, the eigenvalues of the Laplacian operators $L^1$ and $L^2$, for two choices of the intra-layer diffusion. When $D_v^{12}=0$ the two dispersion relations (circles, respectively red and blue online), each associated to one of the independent layers, are negative as they both fall below the horizontal dashed line. For $D_v^{12}=0.5$ the curves lift, while preserving almost unaltered their characteristic profile (square, green online). In particular, the upper branch of the multiplex dispersion relation takes positive values within a bounded domain in $\Lambda_\alpha$, so implying the instability. To confirm the validity of the theoretical predictions we integrated numerically the reaction-diffusion system~\eqref{eq:reac_diff}, assuming the Brusselator reaction terms, and for a choice of the parameters that yield the multiplex instability exemplified in the main plot of Figure~\ref{fig:Fig1}. As expected, the homogeneous fixed point {(dashed line)} gets destabilized: the external  perturbation imposed at time zero, is self-consistently amplified and yields the asymptotic patterns displayed in lower inset of Figure~\ref{fig:Fig1}. 

Interestingly, the dual scenario is also possible. Assign the parameters so that patterns can
develop (on at least one of the layers), in the decoupled setting $D_v^{12}=0$. Then, by increasing $D_v^{12}$, one can eventually remove the instability, and so the patterns, by turning the homogeneous fixed point stable to inhomogeneous external perturbation. Also in this case (demonstrated in the Supplementary Material section), the perturbative theory provides an accurate estimate of the critical value of the intra-layer diffusion constant {(See Figure~\ref{fig:FigApp} SM)}.

Summing up we have developed a consistent theory of patterns formation for a reaction diffusion system defined {upon} a stratified multiplex network. The analysis has been here carried out for a two species model, defined on a two layers multiplex. The methodology employed, as well as our main conclusions, readily extend to the general framework where $s$ species are mutually interacting, while diffusing across a $K$ levels multiplex {whose layers can have arbitrary network topologies}. The interference among layers can instigate collective patterns, which are instead lacking in the corresponding uncoupled scenario. Patterns can also evaporate due to the couplings among distinct levels. Conditions for the critical strenght of the coupling constant are given and tested by direct numerical inspection. The hierarchical organization of the embedding space plays therefore a role of paramount importance, so far unappreciated, in seeding the patterns that we see in nature. It is also worth emphasising that novel control strategies could be in principle devised which exploit the mechanisms here characterized.  These potentially interest a large plethora of key applications, which range from the control of the epidemic spreading, to the prevention of the failure of electric networks,  passing through wildlife habitat restorations.  
 
\section*{Acknowledgments}
The work of T.C. presents research results of the Belgian Network DYSCO (Dynamical Systems, Control, and Optimization), funded by the Interuniversity Attraction Poles Programme, initiated by the Belgian State, Science Policy Office. The scientific responsibility rests with its author(s).
D.F. acknowledges financial support of the program Prin 2012 financed by the Italian Miur.

\appendix
\section{Details on the analytical derivation.}

Eq.~\eqref{eqCLambda} contains two unknowns, namely $C_l$ and $\Lambda_l$. To obtain the close analytical solution which is reported in the main body of the paper we observe that Eq.~\eqref{eqCLambda} can be cast in the compact form
\begin{equation}
\label{eq:operator}
[\Lambda_0,X]=Y\, ,
\end{equation}
where $X$ and $Y$ are $\Omega\times\Omega$ matrices and $[\cdot,\cdot]$ stands for the matrix commutator. In practice, given $Y\in \mathbb{R}^{\Omega\times\Omega}$,  one needs to find $X\in \mathbb{R}^{\Omega\times\Omega}$ solution of~\eqref{eq:operator}. Since 
$\Lambda_0$ is a diagonal matrix, the codomain of the operator $[\Lambda_0,\cdot]$ is formed by all the matrices with zero diagonal. To self-consistently solve~\eqref{eq:operator} it is therefore necessary to impose that $Y$ has zero diagonal elements. Hence, matrix $X$ will have its diagonal elements undetermined. 

Because of the above remark one can solve Eq.~\eqref{eqCLambda} by setting $\Lambda_l$ so to cancel the diagonal terms on its right hand side, that is:
\begin{equation}
\label{eq:fixlambda1}
(\Lambda_l)_{ij} = \begin{cases}
(U_0 A_1 V_{l-1})_{ii}-\sum_{k=1}^{l-1}(C_{l-k}\Lambda_k)_{ii} & \text{if $i=j$}\\
0& \text{otherwise}\,. 
\end{cases}
\end{equation}
Then $C_l$ is readily found to match:
\begin{equation}
\label{eq:fixB}
(C_l)_{ij} = \begin{cases}
\frac{(-U_0 A_1 V_{l-1})_{ij}+\sum_{k=1}^{l-1}(C_{l-k}\Lambda_k)_{ij}}{\lambda^{(0)}_i-\lambda^{(0)}_j} & \text{if $i\neq j$}\\
0& \text{otherwise}\, .
\end{cases}
\end{equation}
This latter epression allows us to simplify~\eqref{eq:fixlambda1}.  In fact:
\begin{equation*}
(C_{l-k}\Lambda_k)_{ii}=\sum_h(C_{l-k})_{ih}(\Lambda_k)_{hi}=0\, ,
\end{equation*}
and thus the approximated eigenvalues are given by
\begin{equation}
\label{eq:fixlambda}
(\Lambda_l)_{ij} = \begin{cases}
(U_0A_1V_{l-1})_{ii} & \text{if $i=j$}\\
0& \text{otherwise}\, ,
\end{cases}
\end{equation}
Observe that the previous formulae take a simpler form for $l=1$ when they reduce to:
\begin{equation}
\label{eq:casel2}
\lambda^{(1)}_i =(U_0 A_1V_{0})_{ii} \quad \text{and}\quad (C_1)_{ij} = -\frac{(U_0A_1V_{0})_{ij}}{\lambda^{(0)}_i-\lambda^{(0)}_j} \quad \text{for $i\neq j$.}
\end{equation}

\section{Interference between layers can dissolve the patterns.}

\begin{figure}[h]
\begin{center}
\hspace*{-1cm}
\includegraphics[scale=0.3]{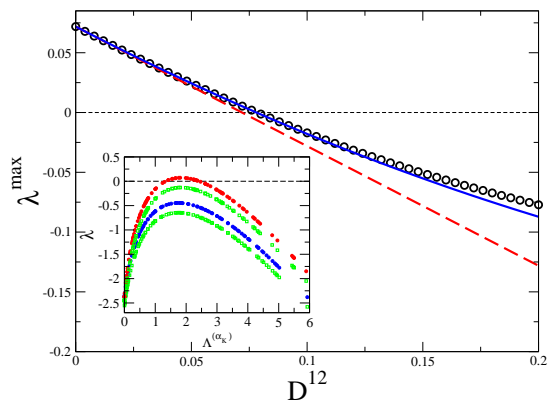}\\
\end{center}
\caption{Main: $\lambda^{max}$ is plotted versus $D^{12} \equiv D_v^{12}=D_u^{12}$, starting from {the value $D^{12}=0$ for which the instability can occur}. Circles refer to a direct numerical {computation} of $\lambda^{max}$. The dashed (resp. solid) line represents the analytical solution as obtained at the first (resp. second) perturbative order. Inset: the dispersion relation $\lambda$  is plotted versus the  eigenvalues of the (single layer) Laplacian operators, $L^1$ and $L^2$. The circles (resp. red and blue online) stand for $D_u^{12}=D_v^{12}=0$, while the squares (green online) are analytically calculated from (\ref{ps}), at the second order, for $D_u^{12}=D_v^{12}=0.2$. The two layers of the multiplex have been generated as Watts-Strogatz (WD) networks with probability of rewiring $p$ respectively equal to  $0.4$ and $0.6$.  The parameters are  $b=8, c=16.2, D^1_u=D^2_u=1, D^1_v=4, D^2_v=5$.}
\label{fig:FigApp}
\end{figure}

We here consider the dual situation as compared to that outlined in  the main body of the paper. We make again reference to the Brussellator model to demonstrate our results. For $\epsilon=0$ the system is unstable, namely $\lambda_0^{max}>0$, as displayed in the main panel of Figure \ref{fig:FigApp}. Patterns can therefore develop on one of the networks that define the multiplex (see unperturbed dispersion relation as plotted in the inset of Figure \ref{fig:FigApp}). The instability is eventually lost for a sufficiently large value of the intra-layer diffusion constant $D^{12}=D_u^{12}=D_v^{12}$. The perturbative calculation that we have developed provides, also in this case, accurate estimates of $\lambda^{max}$ as function of $D^{12}$. The two branches of the dispersion relation shift downward as shown in the inset of {Figure}~\ref{fig:FigApp}.

\label{sec:annex1}


\begin{thebibliography}{99}
\bibitem{murray} J.D. Murray, Mathematical Biology, Second Edition, Springer
\bibitem{zhab} A. M. Zhabotinsky, M. Dolnik and I. R. Epstein, J. Chem. Phys. \textbf{103}, 10306 (1995).
\bibitem{turing} A. M. Turing, Phil. Trans. R. Soc. Lond. B \textbf{237}, 37 (1952).
\bibitem{othmer} H. G. Othmer and L. E. J. Scriven, Theor. Biol. \textbf{32}, 507-537 (1971).
H. G. Othmer and L. E. J. Scriven,  J. Theor. Biol. \textbf{43}, 83-112 (1974).
\bibitem{nakao} H. Nakao and A. S. Mikhailov, Nature Physics \textbf{6}, 544 (2010).
\bibitem{asllaniNatureComm} M. Asllani et al. Nature Communications to appear (2014).
\bibitem{multi1} P.J. Mucha et al. Science \textbf{328} 876 (2010).
\bibitem{multi2} J. Gomez-Gardenes, I. Reinares, A. Arenas. L.M. Floria, Scientific Reports\textbf{2} 620 (2012)
\bibitem{multi3} G. Bianconi, Phys. Rev. E \textbf{87} 062806 (2013).
\bibitem{multi4} R.G. Morris and M. Barthelemy, Phys. Rev. Lett. \textbf{109}
 128703 (2012)
.\bibitem{multi5} V. Nicosia, G. Bianconi, V. Latora and M. Barthelemy, Phys. Rev. Lett. \textbf{111} 058701 (2013) 
\bibitem{kuran} M. Kurant, P. Thiran, Phys. Rev. Lett. \textbf{96} 138701 (2006).
\bibitem{zou} S. R. Zou, T. Zhou, A. F. Liu, X. L. Xu, and D.R. He,
Phys. Lett. A \textbf{374} 4406 (2010).
\bibitem{brain} E. Bullmore and O. Sporns, Nat. Rev. Neurosci. \textbf{10}
186 (2009)
\bibitem{social} S. Wasserman and K. Faust, Social Network Analysis:
Methods and Applications (Cambridge University Press, Cambridge, England, 1994), Vol. 8
\bibitem{arenas} S. Gomez et al Phys. Rev. Lett. \textbf{110} 028701 (2013). 
\bibitem{golub} G. Golub and Ch. F. Van Loan,
\newblock {\em Matrix computations}
\newblock 3rd edition, the Johns Hopkins University Press, Baltimore (Maryland) (1996) 
\bibitem{WS} D. J. Watts, S. H. Strogatz, Nature \textbf{393} 440-442 (1998).
\end{thebibliography}
\end{document}